# An extended halo around an ancient dwarf galaxy

Anirudh Chiti[1*], Anna Frebel[1], Joshua D. Simon[2], Denis Erkal[3], Laura J. Chang[4], Lina Necib[2], Alexander P. Ji[2,5], Helmut Jerjen[6], Dongwon Kim[7], John E. Norris[6]

The Milky Way is surrounded by dozens of ultra-faint (<$10^5$ solar luminosities) dwarf satellite galaxies[1,2,3]. They are the surviving remnants of the earliest galaxies[4], as confirmed by their ancient (~13 billion years old)[5] and chemically primitive[6,7] stars. Simulations[8,9,10] suggest that these systems formed within extended dark matter halos and experienced early galaxy mergers and supernova feedback. However, the signatures of these events would lie outside their core regions (>2 half-light radii)[11], which are spectroscopically unstudied due to the sparseness of their distant stars[12]. Here we identify members of the Tucana II ultra-faint dwarf galaxy in its outer region (up to 9 half-light radii), demonstrating the system to be dramatically more spatially extended and chemically primitive than previously found. These distant stars are extremely metal-poor (<[Fe/H]>=−3.02; less than ~1/1000th of the solar iron abundance), affirming Tucana II as the most metal-poor known galaxy. We observationally establish, for the first time, an extended dark matter halo surrounding an ultra-faint dwarf galaxy out to one kiloparsec, with a total mass of >$10^7$ solar masses. This measurement is consistent with the expected ~$2 \times 10^7$ solar masses using a generalized NFW density profile. The extended nature of Tucana II suggests that it may have undergone strong bursty feedback or been the product of an early galactic merger [10,11]. We demonstrate that spatially extended stellar populations, which other ultra-faint dwarfs hint at hosting as well[13,14], are observable in principle and open the possibility for detailed studies of the stellar halos of relic galaxies.

Tucana II is a typical ultra-faint dwarf galaxy: it is extremely dark-matter-dominated (M/L~2000[15]), has a low metallicity (<[Fe/H]> = −2.7[16,17]), and has a low stellar mass (~3000 solar masses[18]). As with other similar systems[3], spectroscopy of its member stars remains sparse due to its low stellar density[18,19]. Previous follow-up spectroscopic studies were largely limited to stars within two half-light radii[15,16,17] and identified ten probable members of Tucana II.

[1] Department of Physics & Kavli Institute for Astrophysics and Space Research, Massachusetts Institute of Technology, Cambridge, MA 02139, USA
[2] Observatories of the Carnegie Institution for Science, Pasadena, CA 91101, USA
[3] Department of Physics, University of Surrey, Guildford GU2 7XH, UK
[4] Department of Physics, Princeton University, Princeton, NJ 08544, USA
[5] Hubble Fellow
[6] Research School of Astronomy and Astrophysics, Australian National University, Canberra, ACT 2611, Australia
[7]  Department of Astronomy, University of California, Berkeley, Berkeley, CA 94720-3411

To significantly extend the spectroscopic characterization of Tucana II, we obtained wide-field images (~2 x 2 degrees) with the ANU 1.3 meter SkyMapper telescope[20] and used its unique filter-set to efficiently identify metal-poor red giant stars at large galactocentric distances[21]. This efficiency arises because the filter-set enables the derivation of stellar metallicity and surface gravity solely from photometry. By combining these derived stellar parameters with Gaia DR2 proper motions [22], we identified new candidate member stars in Tucana II in a spatially unbiased manner. We then verified their membership and spectroscopically characterized nine of these stars, nearly doubling the previously known stellar population of this galaxy. These stars were detected out to ~9 times the half-light radius[18] (~1 kiloparsec) of Tucana II -- the first detection of member stars beyond ~4 half-light radii in any ultra-faint dwarf galaxy.

Our follow-up spectroscopic observations of candidate members were performed with the MagE[23] and IMACS[24] instruments on the 6.5 meter Magellan-Baade telescope. These spectra enable measurement of radial velocities with a precision of ~3 km/s and ~1 km/s, respectively, and metallicities with a precision of ~0.2 to ~0.3 dex. Such precisions are sufficient to conclusively determine the membership status of all candidate stars from a joint velocity and metallicity analysis.

The metallicities of the spatially extended members decrease the Tucana II galactic metallicity to <[Fe/H]>=−2.77, affirming Tucana II as the most metal-poor galaxy known. Metallicities from prior work[16,17] already show Tucana II to have a low average metallicity of <[Fe/H]>=−2.71. However, we find that the stars beyond two half-light radii are preferentially more metal-poor (<[Fe/H]>=−3.02) than the already studied core population (see Figure 1), which has a mean metallicity of (<[Fe/H]>=−2.62). Such metallicity gradients have previously been seen in larger dwarf galaxies and are hypothesized to result from, e.g., chemical evolution, feedback, or mergers [25]. Our finding is the first evidence of such a metallicity gradient in a relic early galaxy, indicating that their formation may have also been shaped by the same processes.

The spatial configuration of Tucana II members -- twelve giants within two half-light radii and seven between two and nine half-light radii -- suggests that this ultra-faint dwarf galaxy's stellar density profile may differ from one typically assumed for such systems. Under the common assumption of an exponential density profile, we would not statistically expect to see seven giant stars beyond ~2 half-light radii (~0.24 kiloparsecs[18]) in a sample of 19 members. However, when assuming a Plummer profile, it is unlikely but still principally plausible (at a 7% level) to identify seven giant stars beyond ~2 half-light radii in a sample of nineteen members. Conclusive results rest on a precise knowledge of the half-light radius, which is currently not well constrained (see Methods). Deeper photometry and more precise structural parameters of Tucana II would thus enable a more robust determination of these potential density profile differences. Any such discrepancies might cause concern regarding the existence of these distant members in Tucana II, but with the possible exception of the most metal-rich star (see Methods), every distant star is unambiguously a member. We expect no false positive classifications among our most metal-poor distant stars because the systemic velocity (−129.1 km/s) and low mean metallicity (<[Fe/H]> ~ -2.77) of Tucana II are well separated from those of foreground Milky Way stars (see Methods).

Tidal disruption is the most obvious process to displace stars to large radii, but that explanation is inconsistent with the orbital parameters of Tucana II (see methods). The location of any predicted Tucana II tidal debris, based on its orbit, would be perpendicular to the position of the most distant newly discovered Tucana II member stars (Figure 1). Furthermore, tidally disrupted systems should display a velocity gradient [26], which is not observed in Tucana II. For instance, the radial velocity of the farthest star, at 9.3 half-light radii, is only ~(1±3) km/s away from the systemic velocity of Tucana II. Therefore, Tucana II is currently not tidally disrupting.

It follows that the farthest star in Tucana II must be gravitationally bound to the system, given that Tucana II is not tidally disrupting and that the probability of falsely identifying a member is negligible (see methods). To be bound, the farthest star must lie within the tidal radius of Tucana II. Thus, the tidal radius of Tucana II must extend beyond 1 kiloparsec, which requires an enclosed total mass within 1 kiloparsec of at least $1.3 \times 10^7$ solar masses (see methods). This mass is a factor of 4 larger than the mass within one half-light radius. Such extended, massive dark matter halos of relic galaxies were predicted[27], but previous mass estimates of ultra-faint dwarf galaxies were limited to those within a few hundred parsecs. Our study confirms that the halo of a relic galaxy contains significant mass out to a large distance (~1 kiloparsec) for the first time. The majority of this extended mass distribution of at least $1.3 \times 10^7$ solar masses must consist of dark matter, given the low stellar mass of Tucana II (~3000 solar masses[18]).

We estimate the mass for this extended dark matter halo by attempting to directly model Tucana II with a generalized NFW dark matter density profile. This mass enclosed within ~9 half-light radii comes to $(2.1 +3.7/-1.2) \times 10^7$ solar masses. At face value, this estimate is in excellent agreement with the mass deduced from assuming that the farthest member star is bound and thus adds further evidence that these distant stars are indeed bound to Tucana II. We note that adopting the highest and lowest plausible velocity anisotropy prescriptions only vary this mass at the ~1sigma level (see methods). We show the corresponding enclosed mass and density profiles of Tucana II in Figure 2. To test whether masses at large radii can be extrapolated from estimates within the half-light radius, we also calculated the NFW density profile solely from previously known members[15] and extrapolated to ~9 half-light radii. This extrapolation results in a consistent enclosed mass as inferred from all member stars, and supports the common practice of extrapolating the masses from within one half-light radius to larger radii to, for instance, compare with theoretical models[27]. We also note that our derived mass within 1kpc of ~$2 \times 10^7$ solar masses is consistent with an overall halo mass of ~$10^8$ Msun, roughly compatible with constraints on the minimum virial halo mass [28].

We find tentative evidence for a metallicity gradient in Tucana II, as our more distant member stars tend to have lower metallicities than those in the galaxy core. If such gradients are actually pervasive amongst other ultra-faint dwarf galaxies, then the dwarf galaxy metallicities derived only from core populations may be biased high. This bias might affect prior studies that place ultra-faint dwarf galaxies on the dwarf galaxy mass-metallicity relation, a key prediction from galaxy formation simulations that is sensitive to mechanisms including supernova yields, feedback, and gas accretion [29]. For instance, lowering the mean metallicity of the most metal-poor ultra-faint dwarf galaxies may increase the number of plausible feedback prescriptions in simulating these systems [29].

One way to form the extended stellar halo component of Tucana II is by heating the system via galaxy mergers or stellar feedback. The former interpretation implies Tucana II to be the product of an early merger, likely that of two primitive galaxies at high redshift ($z > ~2$) [8]. Simulations do indeed suggest that a dwarf galaxy with the stellar mass of Tucana II (~3,000 solar masses[18]) should be assembled by no more than a handful of star forming progenitors [30]. Otherwise, early supernova feedback may have heated the most metal-poor stars, which is plausible since ultra-faint dwarf galaxies may have a bursty star formation history [10].

Our detection of a population of stars out to ~9 half-light radii in Tucana II suggests that other ultra-faint dwarf galaxies could plausibly host member stars at large radii as well. Indeed, the ultra-faint dwarf galaxies Segue 1 and Bootes I presently each have one known member star at ~4 half-light radii[13,14]. With targeted wide-field photometric searches, it should be feasible to rapidly uncover the distant members of Segue 1, Bootes I, and additional dwarf galaxies to comprehensively establish the evolution of these early relic systems.

**Corresponding author** Correspondence and requests for material to Anirudh Chiti

**Acknowledgements** Our data were gathered using the 6.5-m Magellan Baade telescope located at Las Campanas Observatory, Chile. A.C. thanks Mattis Magg, Alar Toomre, and Tracy Slatyer for helpful discussions. J.D.S is supported by NSF grant AST-1714873. A.P.J. is supported by NASA through Hubble Fellowship Grant HST-HF2-51393.001, awarded by the Space Telescope Science Institute, which is operated by the Association of Universities for Research in Astronomy, Inc., for NASA, under contract NAS5-26555. H.J. acknowledges support from the Australian Research Council through the Discovery Project DP150100862. This work made use of NASAs Astrophysics Data System Bibliographic Services, the SIMBAD database, operated at CDS, Strasbourg, France [31] and the open-source python libraries numpy, scipy, matplotlib, and astropy.

This project used public archival data from the Dark Energy Survey (DES). Funding for the DES Projects has been provided by the U.S. Department of Energy, the U.S. National Science Foundation, the Ministry of Science and Education of Spain, the Science and Technology Facilities Council of the United Kingdom, the Higher Education Funding Council for England, the National Center for Supercomputing Applications at the University of Illinois at Urbana–Champaign, the Kavli Institute of Cosmological Physics at the University of Chicago, the Center for Cosmology and Astro-Particle Physics at the Ohio State University, the Mitchell Institute for Fundamental Physics and Astronomy at Texas A&M University, Financiadora de Estudos e Projetos, Fundação Carlos Chagas Filho de Amparo à Pesquisa do Estado do Rio de Janeiro, Conselho Nacional de Desenvolvimento Científico e Tecnológico and the Ministério da Ciência, Tecnologia e Inovação, the Deutsche Forschungsgemeinschaft and the Collaborating Institutions in the Dark Energy Survey.

The Collaborating Institutions are Argonne National Laboratory, the University of California at Santa Cruz, the University of Cambridge, Centro de Investigaciones Enérgeticas, Medioambientales y Tecnológicas–Madrid, the University of Chicago, University College London, the DES-Brazil Consortium, the University of Edinburgh, the Eidgenössische Technische Hochschule (ETH) Zürich, Fermi National Accelerator Laboratory, the University of Illinois at Urbana-Champaign, the Institut de Ciències de l'Espai (IEEC/CSIC), the Institut de Física d'Altes Energies, Lawrence Berkeley National Laboratory, the Ludwig-Maximilians Universität München and the associated Excellence Cluster Universe, the University of Michigan, the National Optical Astronomy Observatory, the University of Nottingham, The Ohio State University, the OzDES Membership Consortium, the University of Pennsylvania, the University of Portsmouth, SLAC National Accelerator Laboratory, Stanford University, the University of Sussex, and Texas A&M University.

Based in part on observations at Cerro Tololo Inter-American Observatory, National Optical Astronomy Observatory, which is operated by the Association of Universities for Research in Astronomy (AURA) under a cooperative agreement with the National Science Foundation.

This work has made use of data from the European Space Agency (ESA) mission Gaia (https://www. cosmos.esa.int/gaia), processed by the Gaia Data Processing and Analysis Consortium (DPAC, https://www. cosmos.esa.int/web/gaia/dpac/consortium). Funding for the DPAC has been provided by national institutions, in particular the institutions participating in the Gaia Multilateral Agreement.

The national facility capability for SkyMapper has been funded through ARC LIEF grant LE130100104 from the Australian Research Council, awarded to the University of Sydney, the Australian National University, Swinburne University of Technology, the University of Queensland, the University of Western Australia, the University of Melbourne, Curtin University of Technology, Monash University and the Australian Astronomical Observatory. SkyMapper is owned and operated by The Australian National University's Research School of Astronomy and Astrophysics.

This research uses services or data provided by the Astro Data Archive at NSF's OIR Lab. NSF's OIR Lab is managed by the Association of Universities for Research in Astronomy (AURA) under a cooperative agreement with the National Science Foundation.

**Author Contributions** A.C. selected candidates for the observations, took the observations, and led the analysis and paper writing; A.F. assisted with the MagE observations and subsequent analysis, and J.D.S. assisted with the IMACS observations and subsequent analysis; H.J., D.K., and J.E.N. provided the SkyMapper images from which targets were selected; D.E. modeled the orbit of Tucana II; L.J.C. and L.N. modeled the extended density profile of Tucana II; A.P.J contributed to the analysis of the MagE spectra; all authors contributed to the interpretation of the data, and contributed to the paper or provided feedback.

**Data Availability** The measurements that support the findings of this study are presented in extended data tables 1, 2, and 3. The individual stellar spectra from which these measurements were derived and any supplementary material (e.g., linelists) are available from the corresponding author upon reasonable request. The proper motions of the stars analysed in this paper are publicly available from the Gaia DR2 archive (http://gea.esac.esa.int/archive/).

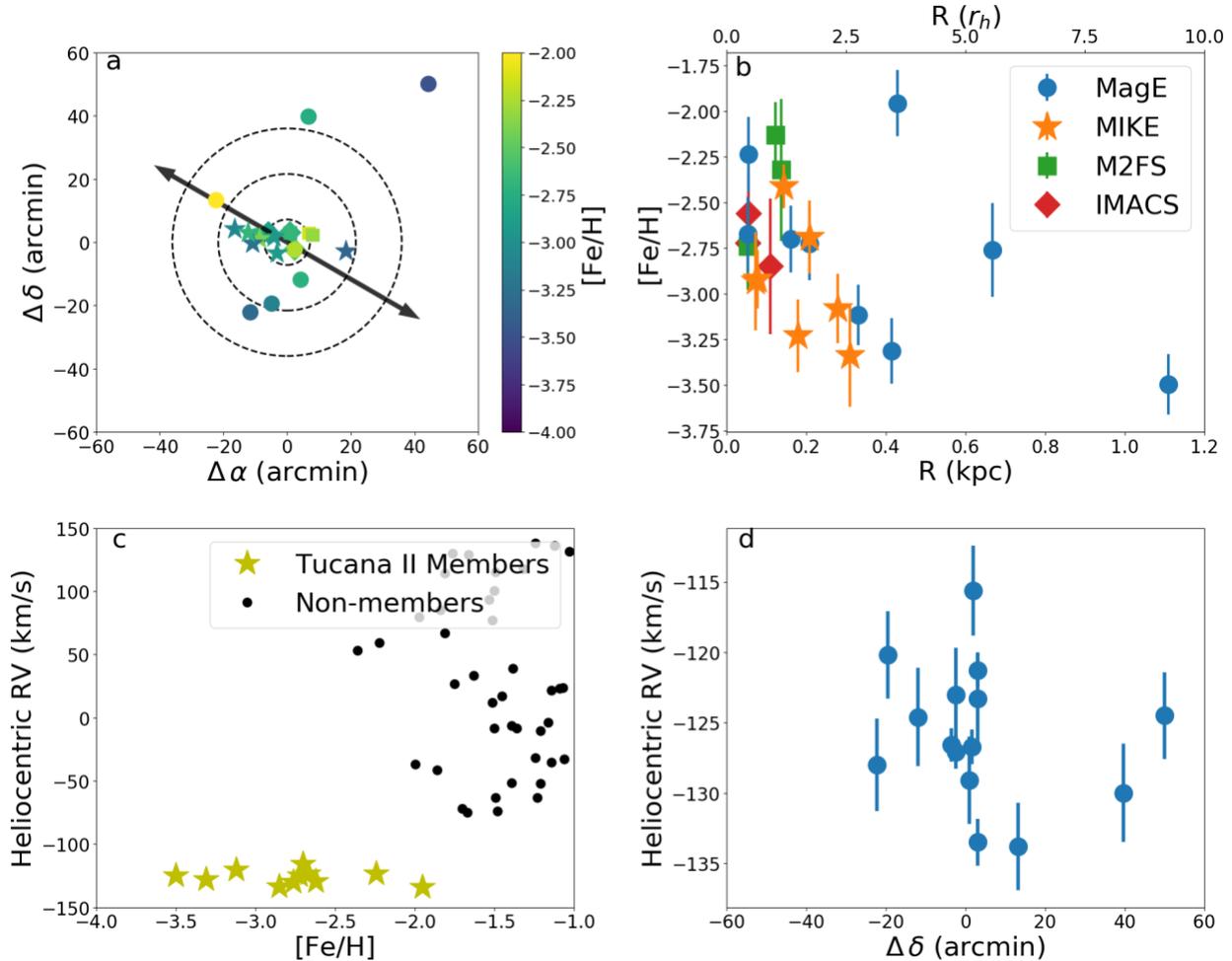

**Figure 1:** Spatial distribution, radial velocities, and metallicities of Tucana II stars as a function of distance from the center of the system
**a.** Spatial distribution of all confirmed member stars of the Tucana II ultra-faint dwarf galaxy, colored by metallicity. The dashed ellipses correspond to one, three, and five half-light radii [19]. Metallicities from MIKE high-resolution spectra are shown as stars [17], those from M2FS spectra are squares [15], and those from MagE and IMACS spectra presented in this work are circles and diamonds, respectively. For Tucana II stars with no high-resolution MIKE results, we plot all our available medium-resolution measurements. Metallicities from M2FS spectra are reduced by 0.32 dex for agreement with high-resolution metallicities (see methods). Contours correspond to 1, 3, and 5 half-light radii. Arrows indicate the direction of predicted Tucana II tidal debris (see methods). Our distant members lie perpendicular to this track, suggesting that their distant location is not due to tidal disruption.

**b.** Metallicities of Tucana II member stars as a function of their geometric radius from the center of the system. There is a general trend towards lower metallicities at larger distances. As for panel (a), metallicities from M2FS spectra are reduced by 0.32 dex. The error bars correspond to 1sigma uncertainties on the metallicity, as derived in the methods section.

**c.** Velocities and metallicities of our IMACS and MagE Tucana II members (yellow stars) compared to non-members in those samples with metallicity measurements and non-members with S/N > 5 observed with M2FS[15] (black points). There is a clear separation in metallicity and velocity space between the two populations.

**d.** Heliocentric radial velocities from MagE and IMACS measurements of Tucana II members as a function of distance (declination) from the center of the system. A velocity gradient would be present in the data if the newly discovered member stars were being dispersed due to tidal stripping. The error bars correspond to 1sigma uncertainties on the velocity measurements, as derived in the methods section.

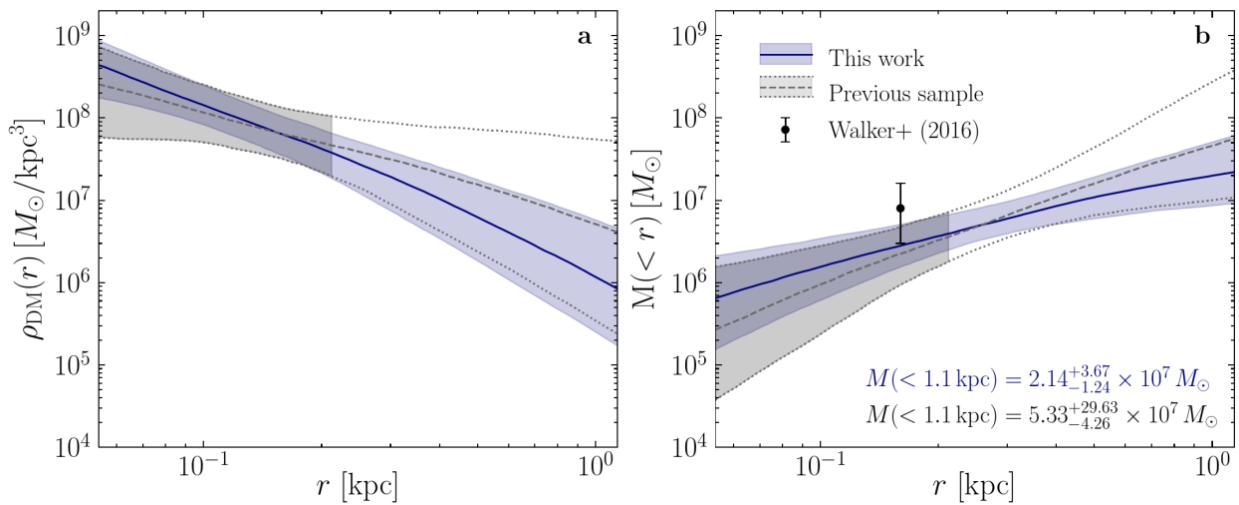

**Figure 2:** Mass modeling of Tucana II out to 1 kiloparsec
**a.** Density profiles of Tucana II, derived using data from previous spectroscopic work [15] shown in grey, and from including the new stars in blue. The error bounds correspond to 1sigma uncertainties from the posterior density distribution.

**b.** Enclosed mass as a function of distance, derived using data from previous spectroscopic studies[15] shown in grey, and from including the new members, in blue. An enclosed mass measurement within a half-light radius[18] using the velocity dispersion and 1 sigma uncertainty from ref[15] is shown as the black data point. The error bounds correspond to 1sigma uncertainties from the posterior mass distribution in our modeling.

| Name | RA | DEC | g | g−i | rv_hel (km/s) | e_rv (km/s) | [Fe/H] | e_[Fe/H] |
|---|---|---|---|---|---|---|---|---|
| Tuc2-301 | 22:50:45.097 | −58:56:20.483 | 18.87 | 0.57 | −128.0 | 3.3 | −3.31 | 0.18 |
| Tuc2-303 | 22:53:05.194 | −57:54:27.032 | 18.44 | 0.54 | −130.0 | 3.5 | −2.76 | 0.25 |
| Tuc2-305 | 22:57:46.859 | −57:43:39.299 | 18.47 | 0.60 | −124.5 | 3.1 | −3.50 | 0.17 |
| Tuc2-306 | 22:51:37.019 | −58:53:37.579 | 18.38 | 0.67 | −120.2 | 3.1 | −3.12 | 0.17 |
| Tuc2-309 | 22:49:24.690 | −58:20:47.429 | 18.73 | 0.63 | −133.8 | 3.1 | −1.96 | 0.18 |
| Tuc2-310 | 22:52:47.376 | −58:46:04.102 | 19.12 | 0.44 | −124.6 | 3.5 | −2.73 | 0.20 |
| Tuc2-318 | 22:51:08.309 | −58:33:08.129 | 18.47 | 0.60 | −129.1 | 3.1 | −2.62 | 0.20 |
| Tuc2-319 | 22:52:32.722 | −58:36:30.488 | 19.33 | 0.58 | −123.0 | 3.3 | −2.24 | 0.21 |
| Tuc2-320 | 22:51:00.921 | −58:32:14.118 | 19.28 | 0.49 | −115.6 | 3.2 | −2.70 | 0.18 |
| Tuc2-321 | 22:52:21.380 | −58:31:07.356 | 19.42 | 0.47 | −123.3 | 3.3 | −2.67 | 0.21 |

**Table 1:** Tucana II member stars observed with MagE.
The Right Ascension (RA) and Declination (DEC) columns indicate the coordinates. g are SkyMapper magnitudes and g−i are extinction-corrected SkyMapper colors. rv_hel lists the heliocentric radial velocity, [Fe/H] the metallicity of each star, together with their associated uncertainties.

## Methods

### Selection of Candidate Members

Targets were initially selected using deep narrow-band imaging of the Tucana II ultra-faint dwarf galaxy obtained using the 1.3m SkyMapper Telescope at Siding Springs Observatory[20] between July and December 2015. The SkyMapper filter-set[32] is unique in that the flux through the *u, v, g*, and *i* filters can be related to stellar metallicity and surface gravity[33,21,34]. Therefore, SkyMapper photometry can directly identify metal-poor red giant stars. These stars are more likely to be members of the Tucana II ultra-faint dwarf galaxy, since the mean metallicity of Tucana II is very low (<[Fe/H]> ~ −2.7[16,17]) compared to most Milky Way halo stars and based on the distance of Tucana II, all members brighter than g~22 should be red giant stars or blue horizontal branch stars[18,19].

In previous work[21], we have shown that one can quantitatively determine stellar metallicities and surface gravities from SkyMapper photometry by generating synthetic photometry[35,36] over a range of stellar parameters and relating observed magnitudes to theoretical magnitudes from this synthetic grid. We applied this method to derive surface gravities and metallicities for every star brighter than g~19.5 in the field of view of the SkyMapper Telescope (~2 degrees x ~2 degrees). We thereby identified many metal-poor ([Fe/H] < −1.0) giant (log g < 3.0) stars as candidate members of Tucana II.

We further refined this sample of likely member stars by making use of Gaia DR2 proper motion data[37, 22]. Since ultra-faint dwarf galaxies are gravitationally bound, their member stars should have proper motions clustered around the systemic proper motion of the galaxy. We narrowed down our sample of candidates by only including stars with proper motions close to the systemic proper motion of Tucana II (pm_ra = 0.936 mas/yr, pm_dec = −1.23 mas/yr [38]). Specifically, we selected stars with proper motions and proper motion uncertainties that are consistent within two sigma of the following bounds:
0.4 mas/yr < pm_ra < 1.2 mas/yr and
−1.4 mas/yr < pm_dec < −1.0 mas/yr (see extended data Figure 1).
These proper motion ranges were chosen to roughly correspond to the proper motions of stars that were previously confirmed to be members of Tucana II[15,17]. Note that we slightly relaxed this proper motion criterion for one candidate member -- it was subsequently found to be a non-member. We then selected a sample of 22 metal-poor red giant stars with proper motions similar to that of Tucana II to observe with the MagE spectrograph. Observing priority was given to stars with photometric [Fe/H] < −1.5, as more metal-poor stars have increased likelihood of being members. Full observational details are given in Extended Data Table 1 below, and the SkyMapper color-color plots from which stars were identified as metal-poor giants are shown in Extended Data Figure 2.

Prior to the *Gaia* DR2 data release and any of our Tucana II SkyMapper photometry studies, we had already taken spectra of 43 stars with the IMACS spectrograph[24] (more details given below). This early target selection was solely based on choosing stars with *g−r* colors within 0.1 mag of a 12 Gyr, [Fe/H] = −2.5 isochrone[39] overlaid at the distance modulus of Tucana II[18,19] on a *g*, *r* color-magnitude diagram of stars within 20 arcmin of the center of the system, along with a few horizontal branch candidates. This color-magnitude diagram had been generated by running the Source Extractor software[40] with default parameters on Dark Energy Survey images of Tucana II obtained from the NOAO public data archive[41,42].

**Spectroscopic Observations**

We obtained spectra of 22 candidate members of the Tucana II ultra-faint dwarf galaxy using the MagE spectrograph[23] on the Magellan-Baade Telescope on 3-5 August 2019. Data were obtained using 1x1 on-chip binning and a 0.7" slit which granted a resolving power of R~6000, and spanning a wavelength range of ~320nm to ~1000nm. The seeing was excellent (<0.6") throughout the duration of these observations. The MagE data were reduced using the CarPy reduction pipeline[43]. From a subsequent radial velocity analysis of these spectra (see Radial Velocity Analysis section below), 10 of these 22 candidates were confirmed as members of Tucana II. A Thorium-Argon lamp frame was obtained for wavelength calibration after slewing to each target. Our targets had *g* magnitudes between 17.5 and 20.5, and each star was observed for at least 10 minutes. For the purposes of more accurate metallicity determinations, the two most metal-poor stars in our sample were observed for more extended periods of time. A full summary of our MagE observations, including total exposure times and signal-to-noise values, is provided in Extended Data Table 1. Examples of our MagE spectra are shown in Extended Data Figure 3.

We had also obtained spectra of 43 stars in the field of the Tucana II ultra-faint dwarf galaxy using the IMACS spectrograph[24] on 23-25 July 2015, 25 May 2016, and 5-7 August 2016. These spectra were obtained simultaneously by using the multi-slit mode of the spectrograph with the f/4 camera, which granted a 15.4 arcmin x 15.4 arcmin field of view. A slit size of 0.7" and a 1,200l/mm grating were used for these observations, which resulted in a resolving power of R~11,000 and a dispersion of 0.19A/pixel over the wavelength range of these spectra (~750nm to ~900nm). We note that this spectral range was chosen to cover the prominent telluric absorption feature at ~760nm and the calcium triplet lines at 849.8nm, 854.2nm, and 866.2nm. Due to the configuration of the multislit mask used for these observations, the actual wavelength range varied from star to star, but slits were placed such that, at minimum, the calcium triplet region was included. Our observing strategy followed that of previous IMACS spectroscopic studies of ultra-faint dwarf galaxy stars[44, 45], except we used a HeNeAr comparison lamp in each of our observations. A total of 14 x 2700s exposures were taken in July 2015 in mediocre seeing conditions (~1.0" seeing), 4 x 2700s exposures in May 2016 in poor seeing conditions ( >1.0"), and 8 x 2700s in generally fair seeing conditions (~0.9") in August 2016. We reduced our data in the exact manner as outlined in the aforementioned studies, utilizing the Cosmos pipeline[46,47] and a modified version of the DEEP2 reduction pipeline[48,49].

**Radial Velocity Analysis**

We derived radial velocities closely following methods by refs[44,45] when analyzing our IMACS spectra, and with slight modifications when analyzing our MagE spectra to account for instrumental differences, differences in wavelength range, and resolution. We briefly describe both analyses here.

The velocities from the IMACS spectra were derived by chi-squared fits[50] to a template IMACS spectrum of the bright, metal-poor giant HD 122563[51] observed in the same configuration as the multi-slit observations. The chi-squared fits were performed over the wavelength range 845nm to 868nm, and a velocity of −26.51 km/s was assumed for HD 122563[52]. Velocity corrections for the mis-centering of stars in their slits were computed by performing this same procedure over the telluric A-band region (750nm to 770nm) using a spectrum of the rapidly rotating hot star HR 4781[44]. These velocity corrections from the A-band region showed a systematic dependence on the position of the slit on the chip[44,45]. Therefore, we fit a linear relation between the velocity correction and the slit position and applied the velocity correction from the A-band following this relation. This had the advantage of being applicable to spectra with A-band regions not covered in the spectral range or with low S/N. Heliocentric corrections were computed using the astropy package[53,54].

The velocities from the MagE spectra were derived by cross-correlating the observed spectra with a template MagE spectrum of HD 122563 observed in the same configuration. The cross-correlation was performed over a wavelength region encompassing the prominent Mg b absorption feature (490nm to 540nm), and the velocity of HD 122563 was again assumed to be −26.51 km/s. The A-band region of an IMACS spectrum of HR 4781[44] that was smoothed to the resolution of the MagE spectra (R~6,700) was used to perform any corrections for the mis-centering of the stars in the slit.

The uncertainties on our velocity measurements were derived by adding in quadrature the statistical uncertainty with an estimate of the systematic uncertainty, following previous work[50,44,45]. To derive the statistical uncertainty, we first re-added random noise to each spectrum based on its noise level as estimated from its signal-to-noise ratio, re-measured the velocity, and then repeated this procedure 500 times[50]. Then, the statistical uncertainty was defined as the standard deviation of the resulting distribution of velocities, after removing 5 sigma outliers. To derive the systematic uncertainty, we computed velocities and statistical uncertainties for spectra that were obtained from individual exposures. We then derived the systematic uncertainty as the additional uncertainty needed to account for the variation in velocities across exposures for each star. The systematic uncertainty for the IMACS spectra was determined to be 1.2 km/s, which agrees with previous work[44, 45]. The systematic uncertainty for the MagE data was determined to be 2.95 km/s, likely due to the lower resolution (R~6,700) relative to the IMACS data (R~11,000).

Final velocities from the IMACS spectra were taken as the weighted average of velocity measurements from stacked spectra from each observing run, where each measurement was weighted by the inverse-square of the uncertainty. We excluded likely binaries from this step, which were identified as stars with at least one two sigma discrepancy in their radial velocities across epochs. Final velocities for the MagE spectra were calculated in the same way, except a weighted average was taken over velocity measurements from stacked spectra for each night of observation. All velocity measurements and uncertainties are listed in Extended Data Table 2.

We derive a systemic velocity for Tucana II of -126.4km/s ± 2.2km/s and a velocity dispersion of 4.6km/s +1.5/-1.1km/s using our sample of MagE and IMACS members. We implemented a maximum likelihood estimate following ref[55] using the emcee python package[56] to derive these values. Our systemic velocity and velocity dispersion are consistent with previous of those quantities for Tucana II[15], but have significantly smaller uncertainties. Our measurements imply that Tucana II has a mass of $(2.4 +1.9/-1.2) \times 10^6$ solar masses within a half-light radius, following ref[57]. We note that performing this analysis for a subsample of only the MagE spectra and another subsample of only the IMACS spectra both return Tucana II systemic velocities apart by ~1.3 km/s, which is well within the statistical uncertainty in the systemic velocity. This suggests no significant systematic offset in the velocities derived from each instrument. Excluding the Tucana II star most likely to be a non-member (TucII-309; see "Membership Confirmation") lowers our velocity dispersion by only ~5%.

**Metallicity analysis**

From the MagE spectra, we derived metallicities using the magnesium b absorption feature (~515nm) and the calcium triplet lines (849.8nm, 854.2nm, and 866.2nm). From the IMACS spectra, we solely used the calcium triplet lines. We used an empirical calibration to derive metallicities from the calcium triplet lines[58], and employed standard spectral synthesis techniques to derive metallicities from the magnesium b region. Our particular implementation of these techniques is comprehensively described in prior work with MagE spectra of dwarf galaxy stars[59], which we summarize here.

Stellar metallicities can be derived from the equivalent widths of the calcium triplet lines in combination with the absolute *V* magnitude of the star[58]. For our MagE spectra, we measured the equivalent widths of the calcium triplet lines by fitting the *Voigt1D* model in the *astropy.modeling* package to each line. The spectra were continuum normalized by iteratively fitting a 3rd order spline after masking absorption features. For fits requiring additional attention due to, e.g., poor estimates of the stellar continuum, equivalent widths were measured using the *splot* task in *IRAF*[60,61]. The absolute *V* magnitude was derived for each star using color transformations from the DES photometric system[18] and the distance modulus of Tucana II[18, 19]. Random uncertainties were derived by re-measuring the equivalent widths after varying the continuum level by 1 sigma according to the S/N of each spectrum, and adopting a systematic uncertainty of 0.17dex[58] for the calcium triplet metallicity calibration. For our IMACS spectra, we fit the calcium triplet lines and derive uncertainties following previous studies of dwarf galaxies with IMACS [45]. We find that our IMACS metallicities agree within 1 sigma with literature metallicities for the two stars that have metallicities from previous spectroscopic work of Tucana II[21].

We measured metallicities from the magnesium b region by fitting synthetic spectra of varying abundances to the observed spectrum. The syntheses and fitting were performed with the Spectroscopy Made Hard software[62] using a 2017 version of the MOOG radiative transfer code[63] that has an updated treatment of scattering[64] and the ATLAS9 model atmospheres[65]. The linelist was compiled from various sources[66,67,68,69] using software provided by C. Sneden (priv. comm.). The effective temperature and surface gravity of each star are required as inputs for the spectral synthesis. Initial stellar parameters were derived by matching the *g−r* colors of the Tucana II stars[70] to those on a [Fe/H] = −2.5, 12 Gyr isochrone[39].
To test this method we also derived stellar parameters in this manner for stars with known stellar parameters[21]. We find on average higher effective temperatures by 120 K and higher log g by 0.33 dex compared to the literature results. We therefore correct our stellar parameter measurements by these values in our analysis. We obtain random uncertainties by noting the variation in [Fe/H] required to encompass most of the noise in the absorption feature. Systematic uncertainties were derived by re-measuring [Fe/H] after varying the effective temperature by 150 K and the surface gravity by 0.3 dex. We note that all of our Tucana II members have metallicity measurements dominated by the systematic uncertainty, due to the relatively high signal-to-noise ratio (S/N > 25) of their spectra. Using these methods, we derive metallicities that agree with literature values for the standard stars CD 38−245 ([Fe/H] = −3.97; literature [Fe/H] = −4.06[71]), CS 22897-052 ([Fe/H] = −3.11; literature [Fe/H] = −3.08[71]), and HD 122563 ([Fe/H] = −2.57; literature [Fe/H] = −2.64[51]) that were also observed by the MagE spectrograph.

We find that our metallicities from the magnesium b synthesis generally agree well with metallicities from the calcium triplet method. The metallicity differences have a mean value of 0.01 dex and a standard deviation of 0.26 dex, suggesting no systematic offset. We note that one distant star (Tuc2-309) has significantly different calcium triplet ([Fe/H] = -1.77) and magnesium b ([Fe/H] = -2.47) metallicities. Upon inspection of its spectrum, this star appears to be genuinely deficient in magnesium, or unusually enhanced in calcium, rendering its overall metallicity somewhat ambiguous. For consistency with our other measurements, we still report its metallicity as the weighted average of the calcium triplet and magnesium b metallicities. Further investigation of systematic offsets from the few stars in common between our MagE, IMACS, and MIKE datasets shows that they have metallicities consistent within 2 sigma uncertainties. The stars in common between the samples are indicated in extended Data Table 2.

The final metallicity measurements were derived by taking the weighted average of the metallicities from the calcium triplet lines and the magnesium b region (for the MagE spectra), or simply from the calcium triplet lines (for the IMACS spectra). Due to reduction issues (e.g., bad sky subtraction), we estimate the equivalent width of the reddest calcium triplet line for two stars (Tuc2-303 and Tuc2-319) by taking it to be 0.62 times the sum of the equivalent widths of the other two calcium triplet lines. The value of 0.62 is the mean of the corresponding ratio of equivalent widths for the other Tucana II members. As an additional quality criterion, we only report metallicities from IMACS spectra with S/N greater than or equal to 10. All the metallicities and uncertainties of Tucana II members are presented in Extended Data Table 3.

The mean metallicity of Tucana II was calculated as the average of the metallicities of its member stars, weighted by the squared inverse of their metallicity uncertainty. If available, metallicities from high-resolution MIKE spectra were assumed as the stellar metallicities. Otherwise, metallicities from medium-resolution MagE and IMACS spectra were used. We used metallicities from M2FS spectra[15] for the two stars in that study listed as likely members (membership probability > 0.90) but that were not later re-observed. We reduced these M2FS metallicities by 0.32 dex, which is necessary to undo an artificial offset and bring the metallicities of the entire M2FS sample in agreement with those from high-resolution spectra[21].

We find some tantalizing evidence of a metallicity gradient in Tucana II (see top right panel of Figure 1). A linear fit to the stellar metallicities as a function of their distance from the center of Tucana II returns a statistically significant slope of -0.87dex/kpc ± 0.29dex/kpc. We note that this result rests on the existence of the farthest star in Tucana II; however, given its low metallicity of [Fe/H] = -3.50 there is little doubt that this star is not a member. For reference, if we were to exclude it, the resulting slope of -0.73dex/kpc ± 0.66dex/kpc would no longer be statistically significant. However, excluding the distant star that has the highest probability of being a foreground star, Tuc2-309 (see Section "Membership confirmation"), results again in a more statistically significant slope of -0.99dex/kpc ± 0.24dex/kpc. This highlights the need for a larger sample of distant members to further validate the existence of any such gradient. At face value, our derived metallicity gradient for Tucana II is -0.11dex/$r_h$, assuming the Tucana II half-light radius in ref[18]. This is comparable to other metallicity gradients seen in larger dwarf spheroidal galaxies [72].

**Membership confirmation**

Members of ultra-faint dwarf galaxies are generally identified through a joint analysis of their metallicities and radial velocities, since ultra-faint dwarf galaxy stars tend to be far more metal-poor than foreground Milky Way stars and have clustered radial velocities. Namely, we identified members as stars with radial velocities within thrice the velocity dispersion of Tucana II around the mean radial velocity of Tucana II (−141 km/s to −110 km/s) and with metallicities [Fe/H] ≲ −2.0. It is unlikely that we excluded members based on the velocity threshold- no stars had radial velocities just beyond these cutoffs. However, one distant star (Tuc2-309) has a metallicity just above this threshold ([Fe/H] = -1.95), but a radial velocity and proper motion still consistent with membership. We therefore regard it as a likely member. Both the mean metallicity (~−2.85) and systemic radial velocity (~−125.6 km/s) of Tucana II are well separated from the corresponding values of the foreground stellar population, granting confidence to this particular scheme of confirming membership status. However, given the distant nature of our newly identified members, we performed an additional check on their membership likelihood.

In extended data Figure 4, we show the predicted halo distribution of radial velocities and metallicities for stars in the vicinity of Tucana II from the Besancon stellar population model[73], after replicating our isochrone, [Fe/H], and logg target selection cuts. We find that 0.4% of these stars satisfy our velocity and metallicity criteria for Tucana II membership. By replicating our isochrone, [Fe/H], and logg cuts but relaxing the proper motion cut on our SkyMapper catalog, we estimate that there are 260 foreground metal-poor giant stars within ~9 half-light radii of Tucana II. If 0.4% of these stars satisfy our membership criteria, this would result in ~1 false positive in our sample of Tucana II members. However, we note that the false positive rate drops off rapidly at lower metallicities. Restricting the range of membership metallicities to [Fe/H] < -2.5 results in a rate of 0%. Thus, while one of our more metal-rich Tucana II members could conceivably be a foreground star, there is a negligible chance that the farthest member is falsely classified, given its very low metallicity of [Fe/H] = -3.50. It is also extremely unlikely that any members beyond two half-light radii are falsely classified, given that their metallicities are all below [Fe/H] = -2.5, except for Tuc2-309.

In Extended Data Table 2, we list the membership status of every star in our sample. We identify stars that meet the above radial velocity and metallicity criteria as members. Some stars in our sample have radial velocity measurements that satisfy the velocity criterion, but do not have metallicities as their spectra have S/N < 10. We identify these stars as likely members. We note that all members have proper motions consistent with the bounds defining the Tucana II stellar membership as listed in Section "Selection of Candidate Members."

**Comparison to Canonical Stellar Density Profiles**

We tested whether the spatial distribution of Tucana II member stars is compatible with either an exponential or a Plummer stellar density profile. Specifically, we drew samples of 19 members 10,000 times from each distribution to test the likelihood of observing 12 stars within 2 half-light radii and 7 stars beyond that distance. We find this happens in 7% of cases under a Plummer profile and in 2% of cases under an exponential profile. Excluding our distant member that is most likely to be a foreground star, Tuc2-309, leads to 15% of cases under a Plummer profile fulfilling our criterion. However, we caution that these numbers are very sensitive to the choice of structural parameters for Tucana II.

We assume a half-light radius of 7.2' for Tucana II[18] in this analysis. We opt for this value over others[19, 74] due to better agreement (within 2 sigma) between the structural parameters in ref [18] and those obtained from deeper follow-up photometry[75,76,77,78,79,80] of seven other dwarf galaxies with such available data. The structural parameters from refs[19, 74] for these seven systems show more scatter when compared to the deeper imaging studies. For completeness, we note that if we were to assume the half-light radii in refs [19, 74], the distances of the farthest two stars would be 4.1 and 6.8 half-light radii, and 3.6 and 6.0 half-light radii, respectively. These radii would lead to substantially more agreement with a Plummer profile. As a consequence, we cannot claim a discrepancy with these canonical profiles until deeper photometry of Tucana II is obtained.

From our Extended Data Figure 1 of candidate members, spectroscopic observations of stars can be regarded as complete down to g~19.5 within the inner region of the galaxy, but likely incomplete in the outer regions (beyond 3 half-light radii). This implies that additional distant members of the galaxy may be discovered in the future. If more stars were known in the extended halo, the underlying stellar density distribution would stray even further from any canonical density profile.

Finally, we note that there are multiple distance measurements of Tucana II in the literature, with all measurements being consistent within 2 kpc [18,19,81]. For consistency with our choice of structural parameters, we adopt a distance of 58 kpc from ref[18] throughout our analysis.

**Systemic Proper Motion of Tucana II**

We derive the systemic proper motion of Tucana II by taking the weighted average of the *Gaia* DR2 proper motions[37,22] of members brighter than *g* = 20 in Extended Data Table 2. Each weight was taken as the inverse-square of the measurement uncertainty. We derive a systemic pm_ra = 0.955 mas/yr ± 0.047 mas/yr and a systemic pm_dec = -1.212 mas/yr ± 0.058 mas/yr.

**Modeling of Tidal Disruption**

In order to determine the expected location of tidal debris from Tucana II, we simulate its tidal disruption and subsequent stream formation using the modified Lagrange Cloud stripping technique[82] which has been updated to include the influence of the Large Magellanic Cloud (LMC)[83]. Including the LMC is crucial since it can deflect tidal debris, leading to a significant misalignment between the progenitor's orbit and tidal debris[83,84]. We use a realistic Milky Way model[85] and the machinery of GALPOT[86] to evaluate accelerations in this potential. Motivated by recent fits to the LMC mass[83], we treat the LMC as a Hernquist profile[87] with a mass of $1.5 \times 10^{11}$ solar masses and a scale radius of 17.13 kpc. We integrate Tucana II backwards for 5 Gyr starting from its present day distance[19], proper motions[88], and radial velocity from this work. The LMC is similarly integrated backwards starting with its present day observables[89,90,91]. We model the progenitor of Tucana II as a Plummer sphere with a conservative mass estimate of 2x10$^6$ solar masses[15] and a scale radius of 100 pc. This produces tidal debris which is well aligned with the orbit. This shows that the debris should be aligned with the track shown in Figure 1 with a small offset of -2±4 degrees on the sky. We find that this alignment is not sensitive to the precise choice of the total mass of Tucana II.

We derived the tidal radius of Tucana II using the galpy[92] python library. Specifically, we instantiated an orbit using the orbital parameters of Tucana II[38] and derived a tidal radius using the *rtide* function[93] under the Milky Way potential MWPotential14. We then derived tidal radii for various masses of Tucana II to determine at what mass the tidal radius encompasses the distance to the farthest member (1.11 kiloparsec). This occurred for a mass of 1.3x10$^7$ solar masses, which we take as the lower limit on the Tucana II mass. We note that our choice of MWPotential14 is consistent with reporting a lower bound on the Tucana II mass. MWPotential14 is a low estimate for the Milky Way mass, and adopting a heavier Milky Way mass would only increase the Tucana II mass required to bound its farthest member.

We also tested whether the spatially extended members of Tucana II display a velocity gradient, which is a signature of tidally disrupting systems[26]. To do this, we fitted a line to our MagE and IMACS velocities, weighted by the inverse square of their velocity uncertainty, as a function of declination from the center of Tucana II. These velocities as a function of their declination from the center of Tucana II are shown in the bottom right panel of Figure 1. We find a slope of 0.06km/s/arcmin ± 0.16km/s/arcmin, suggesting that our velocities show no strong evidence of a velocity gradient. Our 2sigma upper limit on the velocity gradient would therefore be 0.32km/s/arcmin. The 2sigma upper limit increases to 0.36km/s/arcmin if Tuc2-309, the distant star most likely to be a non-member, is excluded.

**Modeling of Dark Matter Density Profile**

We model the density profile of the dark matter halo of Tucana II using a Jeans modeling procedure[94,95,96]. For completeness, we briefly outline the most important steps here but more details will be reported in Chang & Necib (in prep).

The three-dimensional distribution of the stars is modeled by a Plummer profile,

$$\nu(r) = \frac{3\,L}{4\,\pi\,a^3}\left(1 + \frac{r^2}{a^2}\right)^{-\frac{5}{2}}$$

where L is the total luminosity and a is the scale length of the distribution. Given that the system is observed in projection, we use the surface brightness profile

$$I(R) = \frac{L}{\pi a^2}(1 + \frac{R^2}{a^2})^{-2}$$

where R is the projected radius, while r is the three dimensional radius from the center of the dwarf. Because the stars contribute negligibly to the gravitational potential, the value of L does not carry physical influence.

We model the density profile of dark matter by a generalized NFW profile[97],

$$\rho_{DM}(r) = \rho_0(\frac{r}{r_s})^{-\gamma}(1 + \frac{r}{r_s})^{-(3-\gamma)}$$

where the free parameters that we fit for are the overall density normalization ($\rho_0$), the scale radius ($r_s$), and the slope of the inner profile ($\gamma$). We assume that the system is spherical and in equilibrium, and solve the projected Jeans equation

$$\sigma_p^2(R) \mid R = 2 \int_R^\infty (1 - \beta(r)\frac{R^2}{r^2})(\frac{\nu(r)\,\sigma_r^2(r)\,r}{\sqrt{r^2 - R^2}})\,dr$$

where $\sigma_p$ is the projected velocity dispersion along the line of sight, $\sigma_r$ is the radial velocity dispersion in 3 dimensions, and $\beta(r) = 1 - (\sigma_\theta^2 + \sigma_\phi^2)/(2\sigma_r^2)$ is the velocity anisotropy of the stars. The radial velocity dispersion is the solution to

$$\frac{1}{\nu}[\frac{\partial}{\partial r}(\nu\,\sigma_r^2) + \frac{2\beta(r)}{r}(\nu\,\sigma_r^2)] = -\frac{G\,M(<r)}{r^2}$$

where G is the gravitational constant, and M(<r) is the enclosed mass within radius r, calculated from the density profile of dark matter.

We use an unbinned likelihood function[98] in order to take into account the individual velocity uncertainties of each star, first assuming that the system is isotropic ($\beta(r) = 0$)). We use a MCMC procedure to find the best fit parameters, assuming uniform priors on Log10(a/kpc) of [-1.2, -0.6], Log10(L/(Lsun)) of [2.0, 4.0], Log10($\rho_0$/(solar masses/kpc^3)) of [2.2, 13.0], Log10[$r_s$/kpc] of [-3.0, 2.0] $\gamma$ of [-1, 3], and average velocity $\bar{v}$/(km/s) of [-200,200].

Given the large swaths of literature on the mass anisotropy degeneracy (see ref[94] and references therein), we also rerun the mass modeling assuming the standard Osipkov-Merritt anisotropy[99,100], defined as

$$\beta(r) = \frac{r^2}{(r^2 + r_a^2)}$$

where $\sigma_r$, $\sigma_\phi$, and $\sigma_\theta$ are the velocity dispersions in spherical coordinates, and $r_a$ is the anisotropy parameter that we fit for, which describes the transition of the anisotropy from beta = 0 at small radii, to beta = 1 at large radii. We assume a prior on $Ln(r_a/kpc)$ of [-5, 0] to ensure the transition radius occurs before the location of the farthest star. We find that the mass of the system with this anisotropy model increases to (4.1 +6.1/-2.68)x $10^7$ solar masses within 1 kpc, which is within a 1sigma variation of the mass in the isotropic case of (2.1 +3.7/-1.2)x$10^7$ solar masses. Furthermore, adopting the most extreme cases of constant velocity anisotropy still only vary our results at the ~1 sigma level relative to the mass derived assuming isotropy. Assuming constant radial anisotropy (beta=1) increases our mass estimate to (7.0 +9.0/-3.0)x$10^7$ solar masses, while assuming extreme constant tangential anisotropy (beta = -9) still results in a large mass of (1.0 +1.6/-0.5)x$10^7$ solar masses within 1 kpc.

As presented in Figure 2, we performed the above analysis on two samples to investigate the effect of adding our newly discovered members to the body known members in the literature. The first sample is simply the previously known red giant members in the literature with precise velocity measurements[15]. The second sample is composed of previously known members and our newly discovered members. We note that the sample of eight red giant members in ref[15] has a systemic velocity (~-127.3 km/s) similar to the systemic velocity of just our newly discovered members (~-126.4 km/s), suggesting no statistically meaningful velocity systematics across the two samples.

We note that the theoretically derived mass estimate from this method, and any larger mass estimates, are supported by our observationally derived lower bound of 1.3 x $10^7$ solar masses that is required to ensure that the most distant Tucana II member is gravitationally bound to the system.

We also derive the enclosed mass within 1 kpc of a $10^8$ solar mass NFW halo to compare our results to recent bounds on the minimum halo mass[28]. We begin this calculation by using the python package Halotools v0.7[101] to initialize a $10^8$ solar mass halo with a concentration parameter of 26 [102]. We then use the *enclosed_mass* method to calculate that a mass of 2.5x$10^7$ solar masses exists within 1 kpc of such a NFW halo, which is comparable to the mass we derive within 1 kpc of Tucana II.

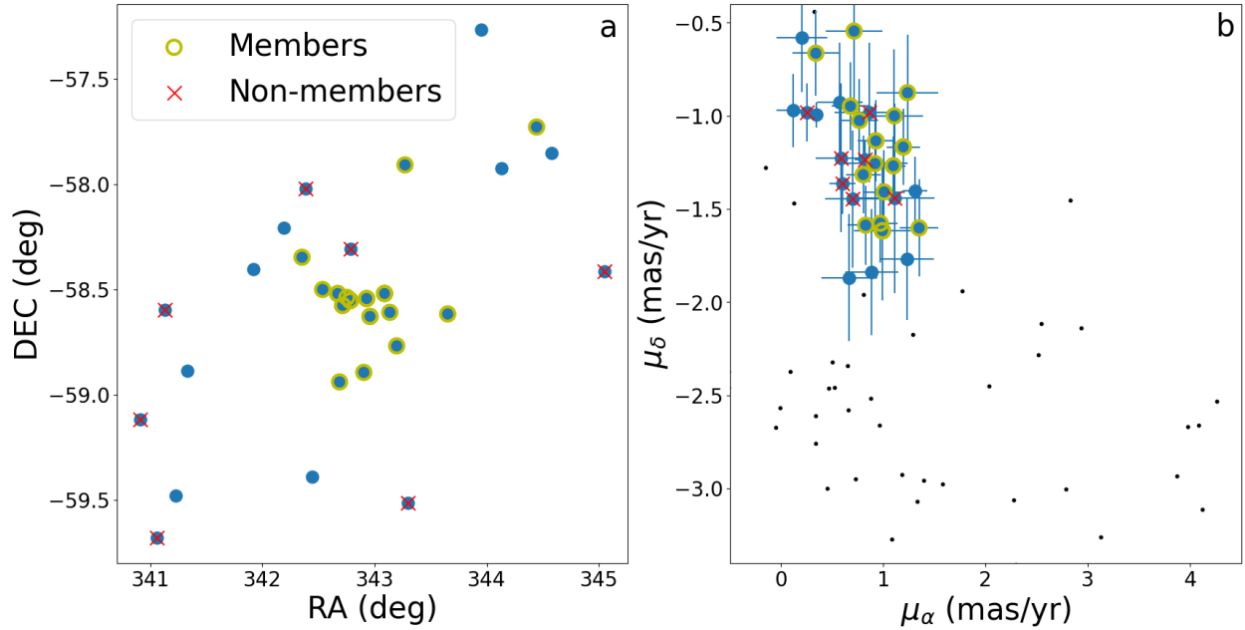

**Extended Data Figure 1** Identification of candidate members of Tucana II
**a.** Locations of candidate members (blue data points) with g < 19.5. Candidates were selected by identifying metal-poor giants with SkyMapper photometry (photometric [Fe/H] < −1.0 and photometric log g < 3.0 [21]), and then only including stars with proper motions around the systemic proper motion of Tucana II (0.2 mas/yr < pm_ra < 1.4 mas/yr and −1.7 mas/yr < pm_dec < −0.5 mas/yr). All stars confirmed as members of Tucana II in this work or prior work [15, 17] are highlighted in yellow. Confirmed non-members of Tucana II that were observed in this work are marked in red.
**b.** Proper motions of candidate members with g < 19.5. The majority of stars with proper motions near the systemic proper motion of Tucana II are members. This results from our exclusion of stars that are not metal-poor giants using log g to cut out foreground stars. Milky Way foreground stars outside our proper motion selection criteria are shown as small black points. The error bars on the proper motions correspond to 1sigma uncertainties in the *Gaia* DR2 catalog.

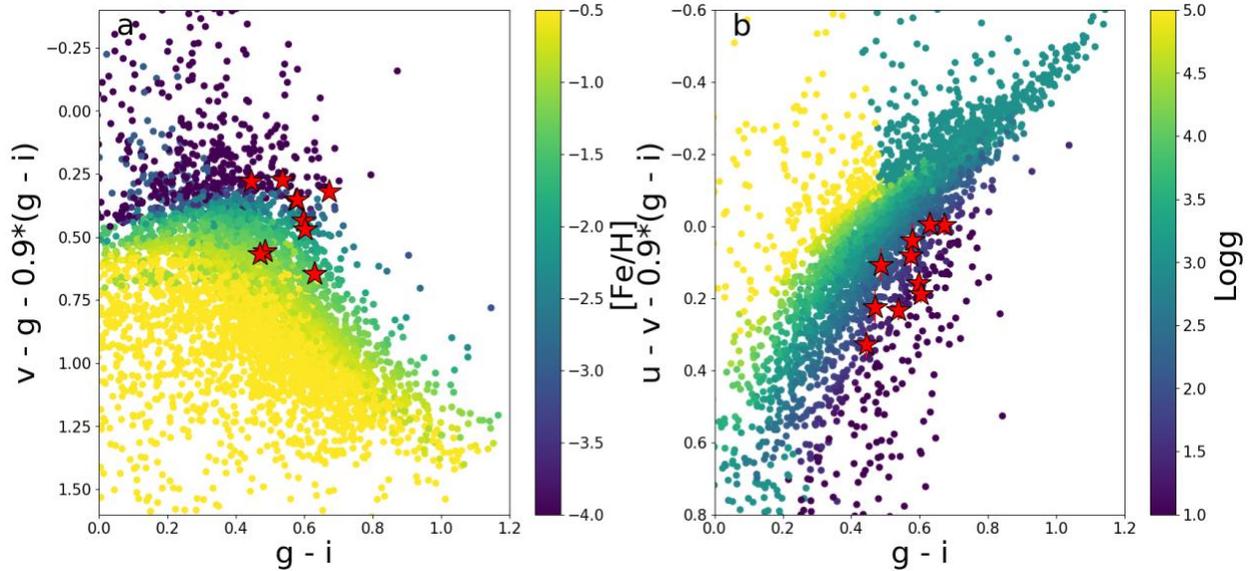

**Extended Data Figure 2:** SkyMapper photometry of Tucana II members observed with MagE

**a.** A metallicity-sensitive SkyMapper color-color plot of every star within a degree of Tucana II. The Tucana II members observed with MagE in this study are shown as purple stars, and all have photometric [Fe/H] < -1.0. Photometric metallicities were derived following ref[21].

**b.** A surface gravity-sensitive SkyMapper color-color plot of every star within a degree of Tucana II. Similarly to the metallicity-sensitive plot, the Tucana II members observed with MagE separate from the foreground population due to their low surface gravities.

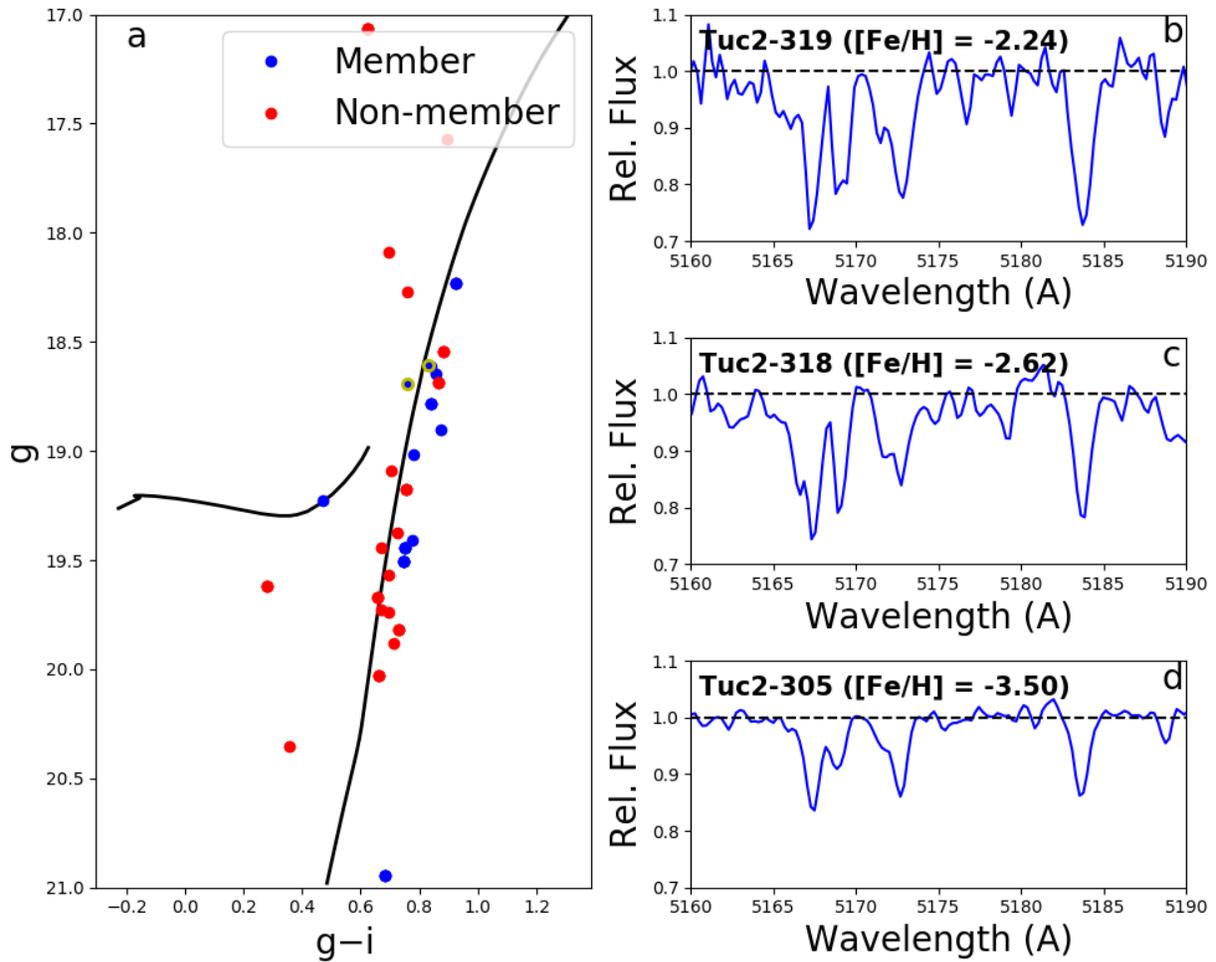

**Extended Data Figure 3** Color-magnitude diagram of Tucana II and sample spectra
**a.** Color-magnitude diagram of the MagE and IMACS Tucana II members with DES photometry. A 10 Gyr, [Fe/H] = -2.2 MIST isochrone[103-107] at the distance modulus of Tucana II[18] is overplotted for reference. The horizontal branch from a PARSEC isochrone[108-113] with the same parameters is also shown. Members and non-members are indicated in blue and red, respectively. The two most distant members are outlined in yellow.
**b,c,d.** MagE spectra of the magnesium region Tuc2-319, Tuc2-318, and Tuc2-305. The absorption lines in the region become noticeably weaker at lower metallicities.

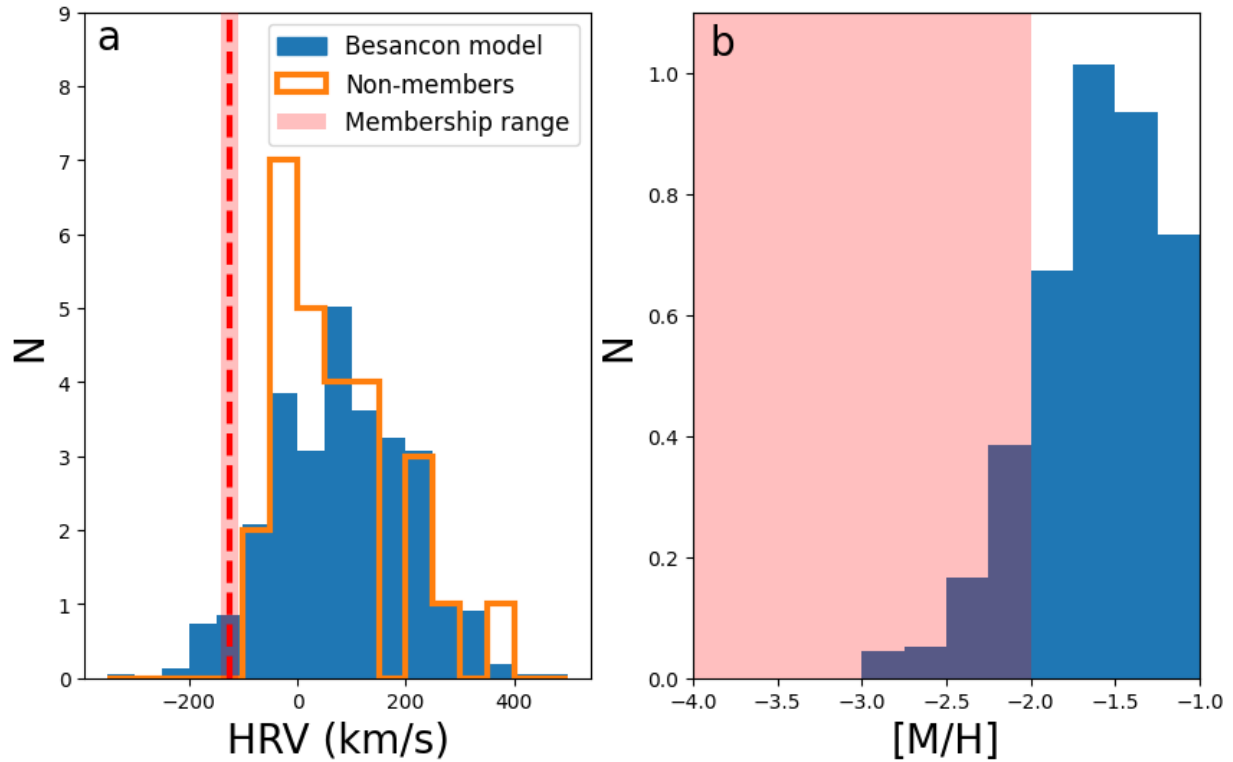

**Extended Data Figure 4** Comparison of Tucana II radial velocities and metallicities to simulated radial velocities and metallicities of foreground stars.

**a.** A histogram of MagE and IMACS radial velocities of stars determined to be non-members of Tucana II is shown in orange. In blue, we plot a scaled histogram of radial velocities of stars in the field of Tucana II, as generated from the Besancon model of stellar populations in the galaxy [73] after replicating our target selection cuts (blue). The vertical red line marks the systemic velocity of Tucana II [15], which is well separated from the foreground velocity distribution.

**b.** Scaled histogram of metallicities of stars generated from the Besancon model following those in panel a. The red shaded region ([Fe/H] < −2.0) corresponds to the metallicities of the newly detected Tucana II members. Only 0.4% of simulated foreground stars satisfy our Tucana II membership criteria (−141 km/s < HRV < −110 km/s; [Fe/H] < −2.0)[15], implying that our newly identified members are extremely unlikely to be false positives.

| Name | RA | DEC | g | t_exp (s) | S/N@500nm |
|---|---|---|---|---|---|
| Tuc2-301 | 22:50:45.097 | −58:56:20.483 | 18.87 | 7800 | 62 |
| Tuc2-302 | 22:49:32.342 | −58:01:12.160 | 19.39 | 2400 | 30 |
| Tuc2-303 | 22:53:05.194 | −57:54:27.032 | 18.44 | 1500 | 30 |
| Tuc2-304 | 22:43:37.613 | −59:07:00.905 | 17.62 | 1500 | 50 |
| Tuc2-305 | 22:57:46.859 | −57:43:39.299 | 18.47 | 10200 | 90 |
| Tuc2-306 | 22:51:37.019 | −58:53:37.579 | 18.38 | 4200 | 55 |
| Tuc2-307 | 22:51:08.442 | −58:18:27.220 | 18.17 | 600 | 12 |
| Tuc2-308 | 22:50:30.170 | −58:20:02.638 | 17.49 | 600 | 36 |
| Tuc2-309 | 22:49:24.690 | −58:20:47.429 | 18.73 | 1800 | 35 |
| Tuc2-310 | 22:52:47.376 | −58:46:04.102 | 19.12 | 2700 | 35 |
| Tuc2-311 | 22:57:58.719 | −59:20:39.127 | 19.55 | 1200 | 15 |
| Tuc2-312 | 23:00:12.426 | −58:24:46.472 | 19.39 | 1200 | 10 |
| Tuc2-313 | 22:59:46.639 | −58:56:02.731 | 20.38 | 2700 | 15 |
| Tuc2-314 | 22:49:20.145 | −57:48:40.845 | 19.67 | 2100 | 20 |
| Tuc2-315 | 22:44:30.549 | −58:35:42.366 | 17.95 | 1500 | 44 |
| Tuc2-316 | 22:53:11.649 | −59:30:56.624 | 19.23 | 1800 | 22 |
| Tuc2-317 | 22:44:14.567 | −59:40:43.309 | 19.01 | 1800 | 29 |
| Tuc2-318 | 22:51:08.309 | −58:33:08.129 | 18.47 | 1500 | 34 |
| Tuc2-319 | 22:52:32.722 | −58:36:30.488 | 19.33 | 1800 | 25 |
| Tuc2-320 | 22:51:00.921 | −58:32:14.118 | 19.28 | 2700 | 30 |
| Tuc2-321 | 22:52:21.380 | −58:31:07.356 | 19.42 | 2700 | 30 |
| Tuc2-322 | 22:57:23.501 | −59:24:00.576 | 19.81 | 1200 | 12 |

**Extended Data Table 1:** Summary of MagE observations

Summary of MagE observations of Tucana II candidate member stars. The Right Ascension (RA) and Declination (DEC) columns indicate the coordinates. g are SkyMapper magnitudes. t_exp lists the total exposure time. S/N@500nm lists the signal-to-noise ratio in the MagE spectrum at 500nm.

**Extended Data Table 2** All velocity measurements and uncertainties *(see table in supplementary material at end of document)*
Summary of velocity measurements of stars in our sample. The MJD column lists the modified Julian date of each observation. The Right Ascension (RA) and Declination (DEC) columns indicate the coordinates. The following column lists DES *g* magnitudes for stars observed with IMACS, and SkyMapper *g* magnitudes for stars observed with MagE. The subsequent column indicates the instrument with which each velocity was determined. The S/N column indicates the signal-to-noise ratio at 850nm. The rv and e_rv columns indicate the velocity measurements and uncertainties. The last column lists membership status.

**Extended Data Table 3** Comprehensive list of metallicities and uncertainties of Tucana II members *(see table in supplementary material at end of document)*
Summary of metallicities of Tucana II member stars. The Right Ascension (RA) and Declination (DEC) indicate the coordinates of each star. [Fe/H]_mg and e_[Fe/H]_mg indicate the metallicities and uncertainties as derived from the magnesium b region at ~515nm. [Fe/H]_CaT and e_[Fe/H]_CaT indicate the metallicities and uncertainties as derived from the calcium triplet region at ~850nm. The instrument with which each metallicity was determined is listed as the last column.

```
Extended Data Table 2
| Gaia DR2 Source ID  | Name    |    MJD  |        RA  |       DEC  |     g  | Instr  |   S/N  |      rv  |  e_rv  | Mem  | Comments
  6503773151318349184 | Star12  | 57228.0 |  342.87278 |  -58.51841 |  20.55 | IMACS  |   7.2  |  -128.3  |   2.9  |  M   |
  6503773151318349184 | Star12  | 57534.4 |  342.87278 |  -58.51841 |  20.55 | IMACS  |   4.8  |  -135.1  |   5.1  |  M   |
  6503773151318349184 | Star12  | 57638.2 |  342.87278 |  -58.51841 |  20.55 | IMACS  |  10.0  |  -134.8  |   1.8  |  M   |
  6503773147023385600 | Star13  | 57228.0 |  342.87738 |  -58.53187 |  18.66 | IMACS  |  39.7  |    85.9  |   1.3  |  NM  | Tuc2-013 in W16
  6503773147023385600 | Star13  | 57534.4 |  342.87738 |  -58.53187 |  18.66 | IMACS  |  18.4  |   147.6  |   1.9  |  NM  | Tuc2-013 in W16
  6503773147023385600 | Star13  | 57638.2 |  342.87738 |  -58.53187 |  18.66 | IMACS  |  39.5  |    92.5  |   1.4  |  NM  | Tuc2-013 in W16
                   -- | Star14  | 57638.2 |  342.88557 |  -58.46953 |  21.58 | IMACS  |   4.2  |   227.7  |   6.0  |  NM  | Tuc2-037 in W16
  6503773559340262144 | Star16  | 57228.0 |  342.90186 |  -58.50471 |  19.29 | IMACS  |  15.5  |  -130.2  |   3.7  |  NM  | Far from CMD
  6503773559340262144 | Star16  | 57534.4 |  342.90186 |  -58.50471 |  19.29 | IMACS  |   8.6  |  -127.8  |  10.1  |  NM  | Far from CMD
  6503773559340262144 | Star16  | 57638.2 |  342.90186 |  -58.50471 |  19.29 | IMACS  |  19.5  |  -115.5  |   1.8  |  NM  | Far from CMD
  6503797340574883072 | Star17  | 57228.0 |  342.93080 |  -58.43178 |  16.21 | IMACS  | 165.3  |    -3.7  |   1.2  |  NM  | Tuc2-054 in W16
  6503797340574883072 | Star17  | 57534.4 |  342.93080 |  -58.43178 |  16.21 | IMACS  |  95.1  |    -2.9  |   1.2  |  NM  | Tuc2-054 in W16
  6503797340574883072 | Star17  | 57638.2 |  342.93080 |  -58.43178 |  16.21 | IMACS  | 141.5  |    -4.3  |   1.2  |  NM  | Tuc2-054 in W16
  6503772322389658880 | Star19  | 57228.0 |  342.92941 |  -58.54270 |  18.22 | IMACS  |  51.6  |  -126.4  |   1.2  |  M   | Tuc2-006 in W16
  6503772322389658880 | Star19  | 57534.4 |  342.92941 |  -58.54270 |  18.22 | IMACS  |  32.3  |  -125.0  |   1.3  |  M   | Tuc2-006 in W16
  6503772322389658880 | Star19  | 57638.2 |  342.92941 |  -58.54270 |  18.22 | IMACS  |  53.5  |  -127.6  |   1.2  |  M   | Tuc2-006 in W16
  6503772704642437760 | Star20  | 57228.0 |  342.94912 |  -58.54708 |  16.61 | IMACS  | 124.5  |    27.0  |   1.2  |  NM  |
  6503772704642437760 | Star20  | 57534.4 |  342.94912 |  -58.54708 |  16.61 | IMACS  |  82.0  |    27.9  |   1.2  |  NM  |
  6503772704642437760 | Star20  | 57638.2 |  342.94912 |  -58.54708 |  16.61 | IMACS  | 116.8  |    25.9  |   1.2  |  NM  |
  6503796275422993920 | Star23  | 57228.0 |  342.97990 |  -58.50262 |  16.99 | IMACS  | 105.5  |    -5.2  |   1.2  |  NM  |
  6503796275422993920 | Star23  | 57534.4 |  342.97990 |  -58.50262 |  16.99 | IMACS  |  53.4  |    -9.6  |   1.2  |  NM  |
  6503796275422993920 | Star23  | 57638.2 |  342.97990 |  -58.50262 |  16.99 | IMACS  |  99.1  |    -6.7  |   1.2  |  NM  |
  6503798208831622240 | Star24  | 57228.0 |  343.01092 |  -58.45373 |  18.08 | IMACS  |  52.3  |    11.4  |   1.2  |  NM  | Tuc2-042 in W16
  6503798208831622240 | Star24  | 57534.4 |  343.01092 |  -58.45373 |  18.08 | IMACS  |  26.9  |    15.2  |   1.5  |  NM  | Tuc2-042 in W16
  6503798208831622240 | Star24  | 57638.2 |  343.01092 |  -58.45373 |  18.08 | IMACS  |  48.7  |    12.5  |   1.3  |  NM  | Tuc2-042 in W16
  6503796580364987392 | Star31  | 57228.0 |  343.07721 |  -58.46650 |  20.50 | IMACS  |   6.0  |   136.0  |   2.4  |  NM  |
  6503796580364987392 | Star31  | 57638.2 |  343.07721 |  -58.46650 |  20.50 | IMACS  |   8.0  |   135.1  |   2.1  |  NM  |
  6491786511775771520 | Star32  | 57228.0 |  343.08909 |  -58.51871 |  19.01 | IMACS  |  24.6  |  -120.8  |   1.3  |  M   | Same as Tuc2-321, Tuc2-022 in W16
  6491786511775771520 | Star32  | 57534.4 |  343.08909 |  -58.51871 |  19.01 | IMACS  |  13.3  |  -120.6  |   2.5  |  M   | Same as Tuc2-321, Tuc2-022 in W16
  6491786511775771520 | Star32  | 57638.2 |  343.08909 |  -58.51871 |  19.01 | IMACS  |  26.9  |  -121.3  |   1.3  |  M   | Same as Tuc2-321, Tuc2-022 in W16
  6503796683444203904 | Star33  | 57228.0 |  343.09579 |  -58.46416 |  20.40 | IMACS  |   5.5  |  -137.9  |   3.0  |  LM  | Likely binary, Tuc2-047 in W16
  6503796683444203904 | Star33  | 57638.2 |  343.09579 |  -58.46416 |  20.40 | IMACS  |   6.7  |  -124.4  |   2.6  |  LM  | Likely binary, Tuc2-047 in W16
  6491786717934204160 | Star38  | 57228.0 |  343.20196 |  -58.51443 |  19.56 | IMACS  |   9.7  |    33.7  |   2.5  |  NM  | Tuc2-049 in W16
  6491786717934204160 | Star38  | 57534.4 |  343.20196 |  -58.51443 |  19.56 | IMACS  |   6.3  |    38.3  |   4.3  |  NM  | Tuc2-049 in W16
  6491786717934204160 | Star38  | 57638.2 |  343.20196 |  -58.51443 |  19.56 | IMACS  |  10.2  |    30.9  |   2.9  |  NM  | Tuc2-049 in W16
  6491786649214722304 | Star40  | 57228.0 |  343.16040 |  -58.52172 |  19.24 | IMACS  |  11.9  |   199.2  |   2.1  |  NM  |
  6491786649214722304 | Star40  | 57534.4 |  343.16040 |  -58.52172 |  19.24 | IMACS  |   7.4  |   240.4  |  11.0  |  NM  |
  6491786649214722304 | Star40  | 57638.2 |  343.16040 |  -58.52172 |  19.24 | IMACS  |  17.4  |   237.8  |   2.3  |  NM  |
                   -- | Star45  | 57638.2 |  342.95030 |  -58.58203 |  21.23 | IMACS  |   4.7  |   398.6  |   5.0  |  NM  |
  6503770024582008320 | Star46  | 57534.4 |  342.92285 |  -58.67116 |  20.61 | IMACS  |   4.2  |   260.2  |   4.6  |  NM  | Tuc2-036 in W16
  6503770024582008320 | Star46  | 57638.2 |  342.92285 |  -58.67116 |  20.61 | IMACS  |   7.8  |   233.0  |   4.8  |  NM  | Tuc2-036 in W16
  6503772189246583936 | Star47  | 57228.0 |  342.91924 |  -58.57088 |  20.90 | IMACS  |   5.2  |   -43.0  |   3.3  |  NM  |
  6503772189246583936 | Star47  | 57638.2 |  342.91924 |  -58.57088 |  20.90 | IMACS  |   6.6  |   -43.6  |   4.7  |  NM  |
                   -- | Star48  | 57638.2 |  342.91436 |  -58.63367 |  21.11 | IMACS  |   4.8  |  -123.7  |   4.3  |  LM  |
  6503772253670168576 | Star51  | 57228.0 |  342.96849 |  -58.56950 |  18.63 | IMACS  |  35.4  |   -34.7  |   1.4  |  NM  | Tuc2-001 in W16
  6503772253670168576 | Star51  | 57534.4 |  342.96849 |  -58.56950 |  18.63 | IMACS  |  22.6  |   -33.2  |   1.5  |  NM  | Tuc2-001 in W16
  6503772253670168576 | Star51  | 57638.2 |  342.96849 |  -58.56950 |  18.63 | IMACS  |  28.0  |   -37.2  |   1.4  |  NM  | Tuc2-001 in W16
  6503770402539968128 | Star55  | 57228.0 |  342.95951 |  -58.62781 |  17.61 | IMACS  |  69.9  |  -126.6  |   1.2  |  M   | Tuc2-011 in W16
  6503770402539968128 | Star55  | 57534.4 |  342.95951 |  -58.62781 |  17.61 | IMACS  |  46.1  |  -127.4  |   1.2  |  M   | Tuc2-011 in W16
  6503770402539968128 | Star55  | 57638.2 |  342.95951 |  -58.62781 |  17.61 | IMACS  |  68.3  |  -126.0  |   1.2  |  M   | Tuc2-011 in W16
  6491762185080920960 | Star57  | 57228.0 |  343.02089 |  -58.63865 |  20.72 | IMACS  |   4.3  |  -115.3  |   5.2  |  LM  | Tuc2-020 in W16
  6491762185080920960 | Star57  | 57638.2 |  343.02089 |  -58.63865 |  20.72 | IMACS  |   7.5  |  -117.6  |   2.4  |  LM  | Tuc2-020 in W16
  6491760638892684544 | Star59  | 57228.0 |  343.04286 |  -58.65961 |  17.92 | IMACS  |  49.3  |   -10.6  |   1.2  |  NM  | Tuc2-027 in W16
  6491760638892684544 | Star59  | 57534.4 |  343.04286 |  -58.65961 |  17.92 | IMACS  |  31.2  |    -9.6  |   1.3  |  NM  | Tuc2-027 in W16
  6491760638892684544 | Star59  | 57638.2 |  343.04286 |  -58.65961 |  17.92 | IMACS  |  52.9  |    -9.6  |   1.2  |  NM  | Tuc2-027 in W16
  6491759608100516224 | Star65  | 57228.0 |  343.11273 |  -58.69541 |  19.32 | IMACS  |  12.9  |   133.7  |   1.7  |  NM  | Tuc2-057 in W16
  6491759608100516224 | Star65  | 57534.4 |  343.11273 |  -58.69541 |  19.32 | IMACS  |   8.1  |   140.2  |   2.1  |  NM  | Tuc2-057 in W16
  6491759608100516224 | Star65  | 57638.2 |  343.11273 |  -58.69541 |  19.32 | IMACS  |  19.3  |   142.3  |   1.8  |  NM  | Tuc2-057 in W16
  6491759606100518016 | Star67  | 57638.2 |  343.12720 |  -58.69269 |  20.98 | IMACS  |   5.3  |   133.1  |   7.5  |  NM  |
  6491762425599106560 | Star68  | 57228.0 |  343.13634 |  -58.60846 |  18.95 | IMACS  |  21.3  |  -128.0  |   1.3  |  M   | Same as Tuc2-319
  6491762425599106560 | Star68  | 57534.4 |  343.13634 |  -58.60846 |  18.95 | IMACS  |  11.7  |  -131.6  |   2.0  |  M   | Same as Tuc2-319
  6491762425599106560 | Star68  | 57638.2 |  343.13634 |  -58.60846 |  18.95 | IMACS  |  26.9  |  -126.5  |   1.4  |  M   | Same as Tuc2-319
  6503385912772220544 | Tuc2-301| 58701.2 |  342.68790 |  -58.93902 |  18.87 | MagE   |  62.0  |  -128.0  |   3.3  |  M   |
  6503821972211079040 | Tuc2-302| 58699.3 |  342.38476 |  -58.02004 |  19.39 | MagE   |  30.0  |   296.2  |   3.4  |  NM  |
  6503910723415846272 | Tuc2-303| 58700.1 |  343.27164 |  -57.90751 |  18.44 | MagE   |  30.0  |  -130.0  |   3.5  |  M   |
  6503333381027472768 | Tuc2-304| 58700.2 |  340.90672 |  -59.11692 |  17.62 | MagE   |  50.0  |   -41.1  |   3.0  |  NM  |
  6493416816937526016 | Tuc2-305| 58701.3 |  344.44525 |  -57.72758 |  18.47 | MagE   |  90.0  |  -124.5  |   3.1  |  M   |
  6491751361762941184 | Tuc2-306| 58700.1 |  342.90425 |  -58.59510 |  18.38 | MagE   |  55.0  |  -120.2  |   3.1  |  M   |
  6503802215362168064 | Tuc2-307| 58701.3 |  342.78517 |  -58.30756 |  18.17 | MagE   |  12.0  |   -38.5  |   3.5  |  NM  |
  6503813592731142400 | Tuc2-308| 58701.3 |  342.62571 |  -58.67870 |  17.49 | MagE   |  36.0  |    85.3  |   3.0  |  NM  |
  6503792319756621184 | Tuc2-309| 58700.3 |  342.35287 |  -58.34651 |  18.73 | MagE   |  35.0  |  -133.8  |   3.1  |  M   |
  6491759092704419840 | Tuc2-310| 58700.2 |  343.19740 |  -58.76781 |  19.12 | MagE   |  35.0  |  -124.6  |   3.5  |  M   |
  6491670753816193408 | Tuc2-311| 58701.3 |  344.49466 |  -59.34420 |  19.55 | MagE   |  15.0  |    13.0  |   3.3  |  NM  |
  6491835062085170816 | Tuc2-312| 58701.3 |  345.05177 |  -58.41291 |  19.39 | MagE   |  10.0  |   113.6  |   3.6  |  NM  |
  6491704220201813376 | Tuc2-313| 58700.3 |  344.94433 |  -58.93409 |  20.38 | MagE   |  15.0  |   -89.7  |   3.5  |  NM  |
  6503877600627638528 | Tuc2-314| 58701.1 |  342.33394 |  -57.81135 |  19.67 | MagE   |  20.0  |   -36.5  |   3.7  |  NM  |
  6503466245840237952 | Tuc2-315| 58701.2 |  341.12729 |  -59.11512 |  17.95 | MagE   |  44.0  |    79.6  |   3.0  |  NM  |
  6491345745051362560 | Tuc2-316| 58701.2 |  343.29854 |  -59.51573 |  19.23 | MagE   |  22.0  |   -52.6  |   3.8  |  NM  |
  6503282146362002176 | Tuc2-317| 58701.2 |  341.06070 |  -59.67870 |  19.01 | MagE   |  29.0  |    39.1  |   3.0  |  NM  |
  6503746932116011280 | Tuc2-318| 58701.2 |  342.78462 |  -58.55226 |  18.47 | MagE   |  34.0  |  -129.1  |   3.1  |  M   | Tuc2-033 in W16
  6491762425599106560 | Tuc2-319| 58701.2 |  343.13634 |  -58.60846 |  19.33 | MagE   |  25.0  |  -123.6  |   3.3  |  M   | Same as Star68
  6503747461931085056 | Tuc2-320| 58699.2 |  342.75384 |  -58.53725 |  19.28 | MagE   |  30.0  |  -115.6  |   3.2  |  M   |
  6491786511775771520 | Tuc2-321| 58699.2 |  343.08908 |  -58.51871 |  19.42 | MagE   |  30.0  |  -123.3  |   3.3  |  M   | Same as Star32, Tuc2-022 in W16
  6491665256258027392 | Tuc2-322| 58701.3 |  344.34792 |  -59.40016 |  19.81 | MagE   |  12.0  |    54.2  |   4.0  |  NM  |
```

Extended Data Table 3

| Name | RA | DEC | [Fe/H]_mg | e_[Fe/H]_mg | [Fe/H]_CaT | e_[Fe/H]_CaT | Instr | Comments |
|---|---|---|---|---|---|---|---|---|
| Tuc2-301 | 342.68790 | -58.93902 | -3.32 | 0.30 | -3.31 | 0.22 | MagE | |
| Tuc2-303 | 343.27164 | -57.90751 | -2.58 | 0.37 | -2.93 | 0.36 | MagE | |
| Tuc2-305 | 344.44525 | -57.72758 | -3.45 | 0.33 | -3.51 | 0.19 | MagE | |
| Tuc2-306 | 342.90425 | -58.89377 | -3.23 | 0.31 | -3.07 | 0.20 | MagE | |
| Tuc2-309 | 342.35287 | -58.34651 | -2.47 | 0.36 | -1.77 | 0.21 | MagE | |
| Tuc2-310 | 343.19740 | -58.76781 | -2.58 | 0.34 | -2.80 | 0.25 | MagE | |
| Tuc2-318 | 342.78462 | -58.55226 | -2.61 | 0.35 | -2.62 | 0.24 | MagE | |
| Tuc2-319 | 343.13634 | -58.60847 | -2.18 | 0.32 | -2.28 | 0.27 | MagE | Same as Star68 |
| Star68 | 343.13634 | -58.60846 | | | -2.56 | 0.22 | IMACS | Same as Tuc2-319 |
| Tuc2-320 | 342.75384 | -58.53725 | -2.75 | 0.35 | -2.68 | 0.22 | MagE | |
| Tuc2-321 | 343.08908 | -58.51871 | -2.63 | 0.30 | -2.71 | 0.28 | MagE | Same as Star32, Tuc2-022 in W16 |
| Star32 | 343.08909 | -58.51871 | | | -2.72 | 0.23 | IMACS | Same as Tuc2-321, Tuc2-022 in W16 |
| Star12 | 342.87278 | -58.51841 | | | -2.85 | 0.37 | IMACS | |
| Star19 | 342.92941 | -58.54270 | | | -3.10 | 0.22 | IMACS | Tuc2-006 in W16 |
| Star55 | 342.95951 | -58.62781 | | | -2.75 | 0.20 | IMACS | Tuc2-011 in W16 |